# Tunable magnetism in 2D organic-ion-intercalated MnPS$_3$ via molecule-dependent vacancy generation

*Daniel Tezze[1], Jose M. Pereira[1], Dogukan Tutar[2], Maria Ramos[1], Jakub Regner[3], Pierluigi Gargiani[4], Frederik Schiller[5,6], Felix Casanova[1,7], Angel Alegria[5,8], Beatriz Martin-Garcia[1], Hasan Sahin[2], Zdenek Sofer[3], Maider Ormaza[8], Luis Hueso[1,7], and Marco Gobbi[5,7]\**

[1] CIC nanoGUNE, 20018 Donostia-San Sebastian, Basque Country, Spain

[2] Department of Photonics, Izmir Institute of Technology, Izmir, Turkey

[3] Dept. of Inorganic Chemistry, University of Chemistry and Technology Prague, Technicka 5, 166 28 Prague 6, Czech Republic

[4] ALBA Synchrotron Light Source, 08290 Barcelona, Spain

[5] Centro de Física de Materiales (CSIC-UPV-EHU) and Materials Physics Center (MPC), 20018 San Sebastián, Spain

E-mail: marco.gobbi@ehu.eus

[6] Donostia International Physics Center (DIPC), 20018 San Sebastián, Spain

[7] IKERBASQUE, Basque Foundation for Science, 48013 Bilbao, Basque Country, Spain

[8] Departamento de Polímeros y Materiales Avanzados: Física, Química y Tecnología (UPV-EHU), 20018 San Sebastián, Spain

The magnetic properties of van der Waals materials are profoundly influenced by structural defects. The layered antiferromagnet MnPS$_3$ offers a unique opportunity to explore defect-related magnetism, as Mn$^{2+}$ vacancies can be generated by the intercalation of specific guest molecules. However, the effectiveness of this process in atomically thin flakes and the extent of the magnetic tunability remain unclear. Here, we show that the magnetic properties of MnPS$_3$ can be tailored through the intercalation of different guest molecules. Notably, the insertion of four alkylammonium ions introduces different populations of Mn$^{2+}$ vacancies, leading to a transition from the pristine antiferromagnetic state to more complex magnetic textures, including a ferrimagnetic state displaying a magnetic saturation of 1 µ$_B$/atom. Moreover, we show that the intercalation of few-nm-thick flakes also leads to the emergence of a ferrimagnetic response. This in-flake intercalation, which can be monitored in real time using optical microscopy, can be interrupted before completion, generating lateral heterostructures between pristine and intercalated areas. This approach opens the way to the use of partial intercalation to define regions with distinct magnetic properties within a single flake.

# 1. Introduction

The discovery of intrinsic ferromagnetic order in atomically thin $CrI_3$[1] and $Cr_2Ge_2Te_6$[2] has sparked significant interest in two-dimensional (2D) layered magnetic materials. These materials represent a highly tunable platform for exploring novel spin-dependent phenomena and designing ultrathin spintronic devices,[3,4] as their magnetic properties can be tailored using electric fields,[5,6] pressure[7] or through the intercalation of guest species.[8–15]

While some materials inherently exhibit magnetic properties,[16] magnetism in other 2D compounds arises from the presence of defects[17–22] or ad-hoc intercalated species.[9,23,24] In particular, numerous studies have revealed that defects in non-magnetic 2D materials generate local magnetic moments, which can acquire (ferro)magnetic ordering.[17–22] Defects can be introduced during the synthesis and exfoliation of the layered materials,[25–27] or through post-growth treatments like thermal annealing, laser irradiation or plasma exposure.[28–31] However, these techniques lack control over the nature, density, and ordering of the defects in the lattice. Additionally, magnetic moments in defect-engineered van der Waals (vdW) materials are typically low, on the order of $10^{-3}$ $\mu_B$/atom.[17]

$MnPS_3$ is a natural layered antiferromagnet[32] which can be exfoliated mechanically[33] or chemically[34], and offers a unique opportunity to explore defect-related magnetism. Specifically, the intercalation of cationic guest species removes $Mn^{2+}$ from the pristine crystalline lattice, inducing vacancies through a cation-exchange process.[35] The resulting intercalated $MnPS_3$ crystals present drastically modified magnetic properties, as the ion-exchange process introduces unpaired spins.[35–38] However, the dependence of the magnetic properties of the intercalates on structural parameters, such as interlayer distance and stoichiometry, has not been thoroughly addressed. As a result, it is not clear to which extent the magnetic properties of $MnPS_3$ can be engineered by controlling the generation of $Mn^{2+}$ vacancies through the use of different molecules. Moreover, while the cation exchange process was explored in bulk crystals four decades ago,[39,40] its effectiveness and feasibility in layered thin flakes is uncertain.

Here, we show that a wide range of magnetic responses can be provided to MnPS$_3$ by mastering the intercalation of four different alkylammonium ions (R$_4$N$^+$). Depending on the specific guest molecule, MnPS$_3$ transits from its pristine antiferromagnetic state to a ferrimagnetic behavior characterized by either a low or a high magnetic moment $M_s$ (as high as 5000 emu/mol, which corresponds to approximately 1 µ$_B$/atom(Mn$^{2+}$)). This vacancy-induced magnetic moment is two-to-three orders of magnitude larger than reported for defect-engineered vdW materials[17] and other intercalated metal phosphorus trichalcogenides.[10] Moreover, we demonstrate that a ferrimagnetic response is also recorded in intercalated micromechanically exfoliated flakes. We optically monitor this in-flake intercalation process, which initiates at the flake edges and proceeds towards the center. The process can be interrupted before the flakes are fully intercalated, enabling the generation of partially intercalated flakes, and potentially offering the means to define domains with different magnetic properties inside the same flake.

## 2. Results and discussion

**Figure 1a, b** shows a sketch of the intercalation method. Bulk MnPS$_3$ crystals (1-2 mg) or exfoliated flakes supported on a Si/SiO$_2$ substrate are immersed in an aqueous solution of alkylammonium bromides salts. The cation exchange process takes place spontaneously, as Mn$^{2+}$ ions are extracted from the MnPS$_3$ lattice by the solvent and the guest organic cations (G$^+$) present in the solution are accommodated in the vdW gap (**Figure 1c**). This process is described by the following heterogeneous equilibrium:

$$2MnPS_{3(s)} + 2xG^+_{(aq.)} \rightleftharpoons 2G_xMn_{(1-\frac{x}{2})}PS_{3(s)} + xMn^{2+}_{(aq.)}$$

We have optimized this one-step intercalation process of bulk MnPS$_3$ crystals for four R$_4$N$^+$ ions with different sizes but similar chemical nature: tetramethylammonium (TMA$^+$), tetraethylammonium (TEA$^+$), tetrabutylammonium (TBA$^+$), and cetyltrimethylammonium (CTA$^+$) (**Figure 1d**, and Table S1, Supporting information). This approach allowed us to obtain highly crystalline monophasic molecular intercalates characterized by a specific stoichiometry, as well as to systematically explore the effect of each cation on the magnetism of MnPS$_3$. We notice that only the TMA$^+$ intercalated MnPS$_3$ had been previously investigated.[38] Unlike the reactions involved in typical electrochemical intercalation processes,[8,41] the ion-exchange mechanism leading to the organic cation intercalation is a non-redox process which does not introduce charge carrier doping in MnPS$_3$. The stoichiometry of the different intercalated MnPS$_3$ crystals was determined by inductively coupled plasma atomic emission spectroscopy (ICP-AES). The ICP-AES measurements lead to the following formula for the intercalates: TMA$_{0.34}$Mn$_{0.83}$PS$_3$, TEA$_{0.34}$Mn$_{0.83}$PS$_3$, TBA$_{0.26}$Mn$_{0.87}$PS$_3$, CTA$_{0.30}$Mn$_{0.85}$PS$_3$ (see Table S2, Supporting Information). We note that these stoichiometries were confirmed by intercalating two sets of crystals, resulting in an error of only 0.01 for the Mn stoichiometric index. Additionally we performed gravimetric and thermal gravimetric analysis (TGA) measurements to confirm the reproducibility of the intercalation and assess the thermal stability of the intercalates (Figure 1 and Table S3, Supporting information). No traces of Br were found in TMA-MnPS$_3$ and TEA-MnPS$_3$, through wavelength dispersive X-ray fluorescence, in agreement with the proposed ion-exchange mechanism (Figure S2, Supporting Information). On the other hand, TBA-MnPS$_3$ and CTA-MnPS$_3$ showed traces of Br, which are likely due to residuals of the salts (TBAB and CTAB) trapped in regions between macroscopic lamellas of the intercalated bulk crystals. These regions, which are not easily accessible to fresh milli-Q water during the washing step, are formed during the intercalation process, due to the expansion of the crystal.

To verify the success of the intercalation process, evaluate the crystallinity, and extract the interlayer distance of each intercalate, X-ray diffraction (XRD) measurements were performed for a pristine $MnPS_3$ bulk crystal and the four $R_4N$-$MnPS_3$ (**Figure 2a** and Table S4, Supporting information). The XRD pattern of the pristine $MnPS_3$ displays a set of sharp (00l) peaks corresponding to a 6.50 Å interlayer distance, in good agreement with the reference pattern (standard JPCDS card PDF#33-0903). Each intercalate displays a single set of sharp (00l) peaks shifted toward lower 2θ values as compared to the pristine material, indicating a larger interlayer distance. These results evidence the success of the intercalation and the formation of single, highly crystalline $R_4N$-$MnPS_3$ superlattices in which $MnPS_3$ monolayers alternate with $R_4N^+$ layers, corresponding to the so-called stage-1 architecture.[42]

In **Figure 2b** we sketch the schematic structure of the obtained superlattices for each of the four hybrid organic-inorganic intercalates, based on the interlayer distance extracted from the XRD patterns using Bragg's law. We notice that TEA-$MnPS_3$ presents a slightly lower interlayer distance than TMA-$MnPS_3$, despite the nominal larger size of the $TEA^+$ cations in solution. We notice that TEA-$MnPS_3$ presents a slightly lower interlayer distance than TMA-$MnPS_3$, apparently in contrast with the one C-atom longer alkyl-chains of $TEA^+$ respect $TMA^+$ molecules. This can be explained considering that $TEA^+$ can adopt two conformational isomers (all-trans conformation, tt.tt or trans–gauche arrangement, tg.tg), characterized by either a spheroidal or a more planar, platelet-like shape. The similar variation of the interlayer distance of TMA-$MnPS_3$ and TEA-$MnPS_3$ (~ 5 Å) is compatible with interlayer $TEA^+$ molecules adopting the trans–gauche (tg.tg) isomerism, laying parallel to the ab-plane. Therefore, the interlayer variation is determined by the tetrahedral N-$C_4$ core, which is structurally similar in the two molecular ions. Finally, in the case of CTA-$MnPS_3$, the interlayer distance is as large as 3.6 nm, indicating that the $CTA^+$ ions are arranged perpendicularly to the $MnPS_3$ planes to maximize the chain-to-chain vdW interaction (see **Figure 2b**).

To further investigate the structure of the R$_4$N-MnPS$_3$ intercalates, we characterized bulk crystals of pristine MnPS$_3$ and of the four bulk intercalates through Raman spectroscopy. **Figure 2c** displays representative Raman spectra between 100 and 650 cm$^{-1}$, which contain the modes corresponding to the vibration of the inorganic MnPS$_3$ layers, while **Figure 2d** shows the 650-1550 cm$^{-1}$ range, which contain the characteristic modes of each alkylammonium species. Importantly, consistent spectra were recorded in different positions of the crystals, indicating a uniform intercalation of the guest species within the host lattice.

Pristine MnPS$_3$ exhibits eight Raman modes (labelled P$_1$-P$_8$), centered at 115, 154, 223, 245, 273, 383, 567 and 580 cm$^{-1}$, respectively (see **Figure 2c**). P$_1$ originates from a cooperative long-range translational mode of Mn$^{2+}$ ions, whereas P$_2$-P$_8$ are related to molecular-like vibrations of the (P$_2$S$_6$)$^{4-}$ bipyramid structural unit.[43] The additional peaks correspond to overtones, with intensities enhanced by the resonance conditions of the excitation source (532 nm laser) with the host material.[44] Upon intercalation of the R$_4$N$^+$ ions, P1 shows a splitting for TMA-MnPS$_3$ and CTA-MnPS$_3$, a considerable attenuation for TEA-MnPS$_3$, and broadening for TBA-MnPS$_3$. Since this mode is associated with the motion of Mn$^{2+}$ ions, these modifications in the intercalated compounds suggest the formation of Mn$^{2+}$ vacancies. P$_2$-P$_8$ remain clearly visible but present a slight shift (see Table S5, Supporting information). These minor changes indicate that the (P$_2$S$_6$)$^{4-}$ units are negligibly distorted but they are arranged slightly differently compared to the pristine material, due to the formation of the Mn$^{2+}$ vacancies and to the presence of R$_4$N$^+$ ions in the interlayer gap.

Moreover, the Raman spectra of the intercalates present two additional features: (i) a well-defined peak is observed at approximately at 626 cm$^{-1}$ for all four intercalates; and (ii) a new peak appears around 136 cm$^{-1}$ for small guest molecules (TMA$^+$ and TEA$^+$), while a broader feature is found in the same range for larger molecules (TBA$^+$ and CTA$^+$). These peaks correspond to IR-active phonons of pristine MnPS$_3$, which become Raman-active after intercalation due to the loss of an inversion symmetry center. [45–47] We attribute this intercalation-induced symmetry breaking again to the creation of Mn$^{2+}$ vacancies, which leads to a long-range distortion of the inorganic lattice previously highlighted.

**Figure 2d** shows the Raman spectra in the 650-1600 cm$^{-1}$ spectral region, in which the pristine MnPS$_3$ material only exhibits weak and broadened overtones. The four R$_4$N$^+$ intercalates still retain traces of these features, confirming the good long-range crystallinity of the lattice. Additionally, the Raman-active modes of the molecular guest species can be detected for each intercalate. A comparison between the Raman spectra taken for the intercalated samples and the solid bromide salts (see Figure S3 and Table S6, Supporting information) reveals shifts for some of their vibrational modes due to the change of the local environment around the molecular ions.[45]

Moving on to the magnetic properties, **Figure 3** shows the magnetic behavior of the pristine MnPS$_3$ and the four alkylammonium-intercalated compounds. Before intercalation, the out-of-plane magnetization vs. temperature curve *M(T)* of bulk MnPS$_3$ (**Figure 3a**) displays the expected Heisenberg-type antiferromagnetic order, characterized by a Néel transition temperature ($T_{Néel}$) of 78 K.[32] On the other hand, a significant change is observed in the *M(T)* for each R$_4$N-MnPS$_3$ intercalate, as a spontaneous magnetization emerges below a specific Curie temperature $T_C$ in the range between 40 - 60 K, and which depends significantly on the intercalated R$_4$N$^+$ guest species (Table S7, Supporting information). All intercalates display ferrimagnetic properties, with low-temperature magnetization at relatively high magnetic fields (5T) varying from approximately $1.0 \times 10^3$ emu/mol for TBA-MnPS$_3$ to $6 \times 10^3$ emu/mol for TMA-MnPS$_3$. Notably, the latter value is 20-100 times larger than that recorded in organic-ion intercalated in NiPS$_3$[10] and vacancies rich-MoS$_2$.[17]

We also note that in the CTA-MnPS$_3$ intercalate, MnPS$_3$ monolayers are separated by 3.6 nm by perpendicularly arranged CTA$^+$ molecules, leading to an absence of coherent interlayer magnetic interactions. In fact, antiferromagnetism in MnPS$_3$ was already demonstrated to persist down to the monolayer limit[33]. Therefore, it is not surprising that magnetism is maintained even in intercalated MnPS$_3$ crystals characterized by a very large interlayer distance, such as CTA-MnPS$_3$. Actually, our data indicate that a variety of magnetic behaviors can be sustained by a MnPS$_3$ monolayer, which can be considered as a tunable and truly-2D magnet.

The importance of the molecular cations in the determination of the magnetic properties of the intercalates becomes evident when comparing the recorded magnetization curves *M(H)* of the four R$_4$N- MnPS$_3$ (**Figure 3b-f**). TMA-MnPS$_3$ shows the highest saturation magnetization M$_S$ ~ 6000 emu/mol (~ 1.1 μ$_B$/atom (Mn$^{2+}$)), and a very narrow hysteresis loop opening (width ~ 40 Oe) with low remanence magnetization *M$_r$* (350 emu/mol out-of-plane, **Figure 3c**). TEA-MnPS$_3$, which has the same stoichiometry and a similar interlayer distance to TMA-MnPS$_3$, shows a much larger coercive field (~ 1250 Oe) with an out-of-plane remanent magnetization of 425 emu/mol, and it does not saturate at 5 T (**Figure 3e**). Importantly, the reproducibility of these magnetic responses was confirmed by the intercalation of another set of different crystals. The results, shown in Figure S4 of the Supporting Information, show a qualitative and quantitative agreement with those reported in Figure 3.

These data indicate that the magnetic response of each intercalate is not solely determined by the number of vacancies or by the interlayer distance, as compounds with nominally identical stoichiometry and similar interlayer distance present very different behavior. In fact, previous studies pointed out the relevance that the ordering of the Mn$^{2+}$ vacancies plays an important role in the determination of the magnetic properties of the intercalates.[35,36] To gain more insight on this point, we perform density functional theory calculations to examine the ground-state magnetic configurations of both pristine MnPS$_3$ and MnPS$_3$ with Mn$^{2+}$ vacancies located in proximity of two TMA$^+$ cations. Our simulations of the pristine MnPS$_3$ crystal structure reveal that Mn$^{2+}$ atoms carry a 5.0 µ$_B$ moment, which are arranged into two antiferromagnetically coupled intralayer hexagonal sublattices, each shifted by one lattice parameter *a*, aligning well with the reported MnPS$_3$ magnetic configuration[32] (**Figure 3g**). The model we used to describe the vacancy-rich TMA-MnPS$_3$ is shown in **Figure 3h**. In this model, one out of six Mn$^{2+}$ atoms is removed from the MnPS$_3$ lattice. This is in agreement with the stoichiometry measured for TMA$_{0.34}$Mn$_{0.83}$PS$_3$, where 17% of Mn$^{2+}$ atoms are removed during the intercalation, or approximately 1 out of 6. We notice that, in our model, the vacancies are orderly arranged, in such a way that they all lie in the same spin sublattice. This scenario was previously proposed for TMA-MnPS$_3$ on the basis of neutron diffraction experiments, which indicated the presence of a magnetic superstructure in the intercalate.[36] Remarkably, our simulations demonstrate that in the presence of TMA$^+$ molecules, the creation of Mn$^{2+}$ vacancies does not alter the overall magnetic arrangement of MnPS$_3$; the two antiferromagnetically coupled spin sublattices remain intact (**Figure 3h**). However, the presence of ordered Mn$^{2+}$ vacancies results in the removal of spins from one sublattice, leaving unpaired spins in the other sublattice. These unpaired spins serve as the source of the net magnetic moment in the intercalated materials.

In this regard, the ordering of the vacancies plays a crucial role in determining the magnetic response of the intercalate. No overall magnetic moment is expected if the Mn$^{2+}$ vacancies are removed randomly from both spin sublattices, whereas a substantial spontaneous magnetization arises when vacancies are preferentially removed from one of the spin sublattices. In the latter case, it is very easy to estimate the expected saturation magnetization as the sum of all unpaired spins. Specifially, if one vacancy is generated for every *N* Mn$^{2+}$ ions in the pristine crystal, the spontaneous magnetization at saturation can be estimated as $M$[Mn$^{2+}$] = 1/($N$-1) × 5.0 µ$_B$/atom (Mn$^{2+}$), where 5.0 µ$_B$ is the magnetic moment of each Mn$^{2+}$ atom in MnPS$_3$. For TMA-MnPS$_3$ (**Figure 3h),** $M$ = 1/5 × 5.0 µ$_B$/atom (Mn$^{2+}$), which is very close to the experimentally measured saturation magnetization for this compound (see Figure 3c). The different magnetic properties recorded in the MnPS$_3$ crystals intercalated with other molecules can be explained on the basis of a different ordering of the vacancies. We put forward toy models for the vacancies distribution to account for the different magnetic responses in Figure S5 of the Supporting Information. In turn, the ordering of vacancies is likely related to the self-assembly of the molecular cations in the vdW gap. Since each molecular guest has a +1 charge and each Mn vacancy has a -2 charge, the vacancy arrangement is determined in an intricate way by the supramolecular forces acting among different molecules, as well as by the Coulomb interaction between vacancies and cations. Therefore, the molecule-dependent magnetic properties can be understood considering that different guest cations form different self-assembled structures, which in turn drive a different ordering of vacancies. However, the experimental characterization of the molecular ordering in the van der Waals gap goes beyond the scope of this work.

Next, we investigate the intercalation process in exfoliated MnPS$_3$ flakes. For this experiment, we focus on the TMA$^+$ intercalation, as it yields the highest saturation magnetization. We first exfoliate MnPS$_3$ flakes from a bulk crystal, and then transfer them on a Si/SiO$_2$ substrate. We use a home-made cell to monitor *in situ* through a microscope the effect of the solution on the flakes (see Methods and Figure S6, Supporting information). We intercalated MnPS$_3$ flakes using a TMAB aqueous solution (1M), as for the intercalation of MnPS$_3$ crystals. **Figure 4a** displays a series of images capturing the intercalation progress of one selected flake, extracted from Supporting Video 1. The selected flake progressively changes color from red in the pristine state to bright green at the end of the process. We ascribe the color change to the interlayer expansion which accompanies the intercalation process. As the different frames show, the TMA$^+$ cations start to occupy the flake from the edges, proceeding towards the center until total intercalation. As a result, an intercalation front is observed which travels from the border to the center at a speed of approximately 2 µm/min for the selected flake. Supporting Video 2 displays another example of in-flake intercalation, showing how flakes of different size and thickness are intercalated in different time scales. We also highlight that TMA$^+$-intercalated flakes exhibit the same interlayer spacing as the bulk crystals and analogous crystalline quality, as confirmed by XRD (Figure S7, Supporting information).

Interestingly, it is possible to interrupt the intercalation process by rinsing the substrate with Milli-Q water. **Figure 4b** shows the optical image of a partially intercalated flake, for which the intercalation was interrupted before completion. Raman spectra were measured to confirm that the different optical contrast corresponds to intercalated and pristine regions. Indeed, the spectra recorded in the outer and inner region of the flake (**Figure 4c**) are characterized by the same features discussed in **Figure 3** for bulk TMA-MnPS$_3$ and pristine MnPS$_3$, respectively. In particular, a clear shift is observed for P$_4$ and P$_8$.

Moreover, we display in **Figure 4d** the spatial variation of the intensity of the P$_5$ and P$_6$ peaks, as well as that of the Si mode of the substrate at 520 cm$^{-1}$. The outer part is characterized by a lower intensity of P$_5$ and P$_6$, probably due to the significant lack of Mn$^{2+}$ ions in the host layer or to a different Raman susceptibility. In contrast, the Si peak in the inner region is more intense than in the outer region, as the reflected light experiences more scattering when travelling across the thicker intercalated region. Overall, the three Raman maps confirm that the outer region is intercalated, while the center is pristine.

Additionally, we imaged the same partially intercalated flake by means of atomic force microscopy (AFM). The fully intercalated outer region is thicker (150 ± 5 nm) than the inner pristine one (90 ± 5 nm), in agreement with the expanded interlayer distance for the TMA$^+$ intercalated MnPS$_3$ bulk crystal (**Fig. 4e**). The two areas are not abruptly separated, as intermediate domains of partially intercalated regions are imaged between them. Also, we notice that the intercalated region is constituted by few-µm large flat domains separated by few-nm-thick wrinkles that form to accommodate the strain introduced by intercalation. The origin of this strain can be attributed to strong in-plane shear forces, generated by the incoming flux of molecules, and to in-plane lattice expansion due to intermolecular repulsion forces. Both phenomena were similarly observed during electrochemical intercalation of $MoS_2$ with $Li^+$ [48,49].

Supplementary Video 2 shows that only flakes thicker than approximately 20 nm (~ 30 layers) exhibit a significant color change upon intercalation. To confirm the successful intercalation of thinner flakes, we have imaged a few-layer-thick flake before and after intercalation by AFM (**Figures 5a** and **b)**. The pristine flake is characterized by a thickness of approximately 6 nm, which corresponds to 6-7 layers. The thickness increases to 9 nm after the process (see **Fig. 5c**), confirming the successful intercalation. Moreover, even for this thin flake, wrinkles are observed after intercalation, similarly to the thicker flake imaged in Figure 4e. This suggests that the intercalation process proceeds in an analogous way for flakes in the range from few layers to more than 100 nm.

To investigate the magnetic properties of intercalated flakes, we conducted X-ray absorption (XAS) and X-ray Magnetic Circular Dichroism (XMCD) experiments at $T$ = 3.5 K. For this experiment, $MnPS_3$ flakes are exfoliated onto a 50-nm-thick Au film deposited on a Si/$SiO_2$ substrate and then intercalated with $TMA^+$ molecules. In order to prevent beam damage, the synchrotron light is defocused to approx. 1x1 $mm^2$ to ensure low photon densities. Therefore, we analyze the magnetic response of a collection of randomly distributed flakes, with thicknesses typically ranging from 10 nm to 100 nm. **Figure 6a** provides an optical image of a representative area of the sample under investigation.

In **Figure 6b**, we present the XAS data recorded at a magnetic field strength of 6 T using both right ($\mu_+$) and left ($\mu_-$) crcularly polarized light using total-electron yield detection. A comparison between the XAS spectra of the TMA$^+$ intercalated and the MnPS$_3$ pristine flakes (Supporting Figure 5) indicates that the organic intercalation does not alter the valence state of the Mn$^{2+}$ ions. The XMCD signal ($\mu_+ - \mu_-$) measured in the TMA$^+$-intercalated MnPS$_3$ is illustrated in the bottom graph of **Figure 6b**. Remarkably, even in the remanent state, i.e., in absence of an applied magnetic field, we observe a discernible XMCD signal (red curve) indicating the presence of a finite magnetic remanence in the intercalated flakes. The behavior of the XMCD signal as a function of the applied magnetic field is shown in the inset of **Figure 6b**. This ferrimagnetic behavior recorded for the intercalated flakes contrasts with the one of the pristine flakes, for which no XMCD signal was observed at zero magnetic field (Figure S8, Supporting information).

In conclusion, we show that the magnetic properties can be manipulated in a controllable way through the intercalation of organic cations, which generate Mn$^{2+}$ vacancies and thus unpaired spins. Notably, when compared to other vdW materials featuring defect-induced magnetism, intercalated MnPS$_3$ exhibits notably larger magnetic moments, on the order of 1 $\mu_B$/atom(Mn$^{2+}$). Moreover, we demonstrate the successful intercalation of few-nm-thick flakes, which also exhibit ferrimagnetic behavior. This process can be monitored in real-time through optical microscopy, which provides a deeper understanding of the intercalation progress and its dynamics. Finally, we created partially intercalated flakes by interrupting the intercalation process before completion, potentially enabling the definition of regions with distinct magnetic properties within a single flake. This novel approach may open doors for the development of lateral heterojunctions with varying physical characteristics defined through partial intercalation of vdW materials, which could find application as semiconducting heterojunctions, spin valves or Josephson junctions.

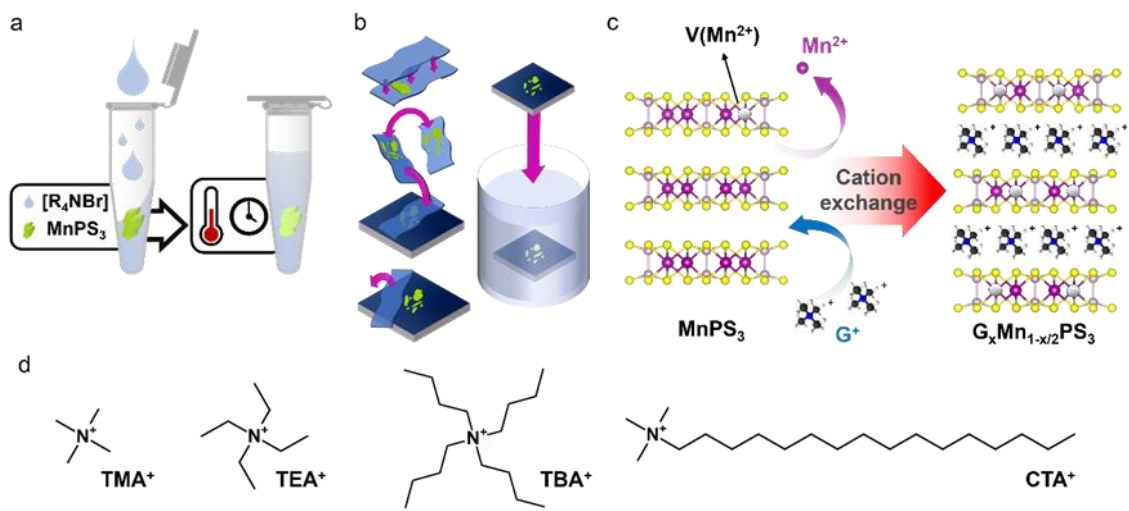

**Figure 1**. Intercalation process and cation-exchange mechanism. a-b) Schematic of the intercalation method of MnPS$_3$ crystals. Bulk MnPS$_3$ crystals (a) or exfoliated flakes on a Si/SiO$_2$ substrate (b) are immersed in an aqueous solution of an alkylammonium bromide salt of a certain concentration, at a needed temperature and for enough amount of time to reach complete intercalation. c) Scheme of the alkylammonium/Mn$^{2+}$ cation exchange mechanism: every two monovalent alkylammonium ions (G$^+$) inserted in the van der Waals gap, a bivalent Mn$^{2+}$ ion (purple spheres) leaves the host MnPS$_3$ lattice, creating an intralayer cationic vacancy (white spheres), V(Mn$^{2+}$). d) Chemical structure of the four alkylammonium ions (TMA$^+$, TEA$^+$, TBA$^+$, CTA$^+$) employed in this work as guest species.

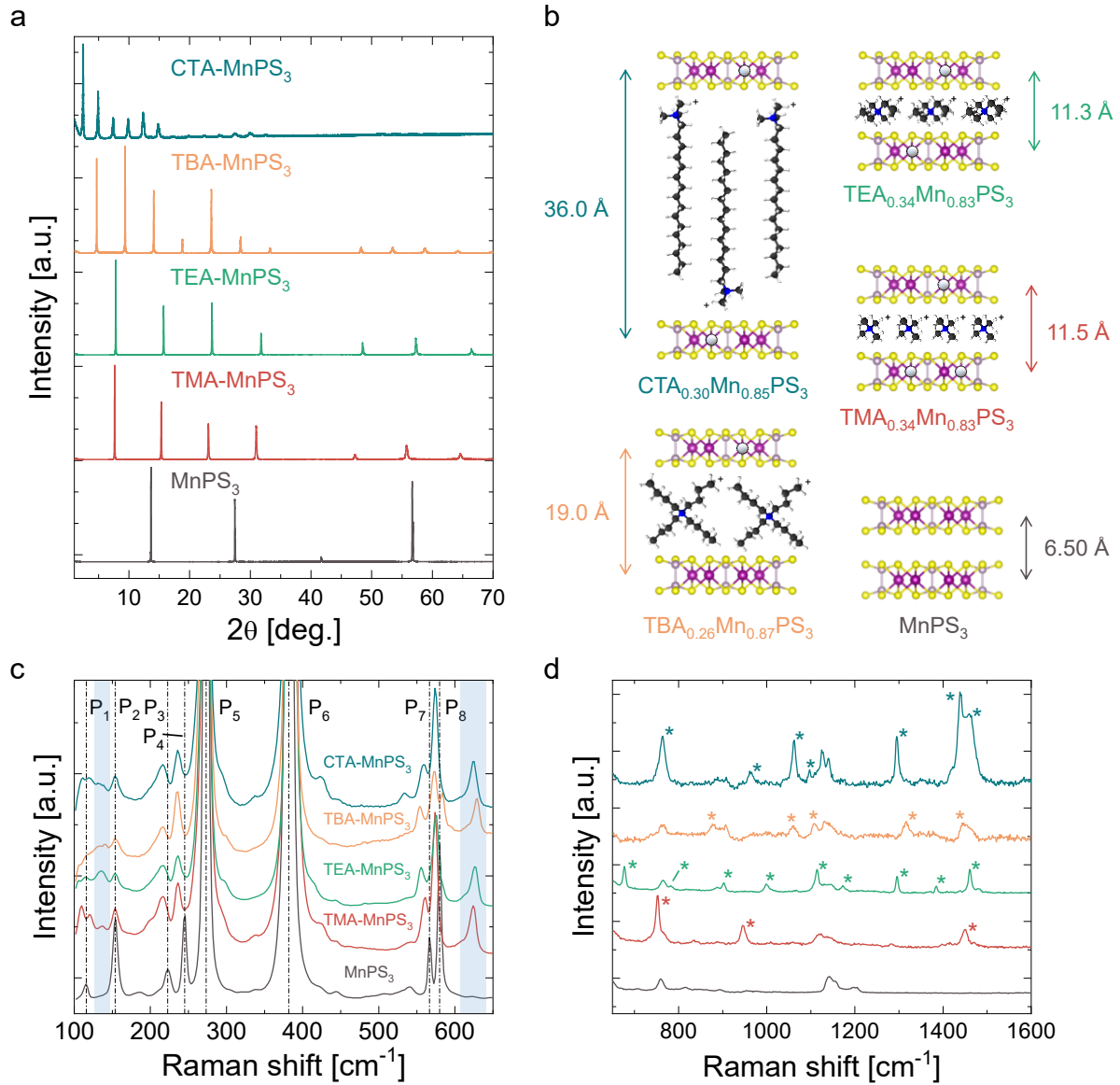

**Figure 2**. Structural characterization of pristine and intercalated bulk crystals of $MnPS_3$. a) X-ray diffraction patterns of pristine $MnPS_3$ and the four $R_4N^+$-$MnPS_3$ intercalates. All the patterns are normalized respect to the intensity of the (001) peak for clarity. b) Schemes of the 2D hybrid organic-inorganic superlattice along the perpendicular direction of the crystal's unit cell, together with their characteristic interlayer distance and stoichiometry. c) Raman spectra in the 100-650 $cm^{-1}$ range and d) in the 650-1600 $cm^{-1}$ range for pristine $MnPS_3$ and the four $R_4N^+$ intercalates. Spectra were acquired at room temperature with a 532-nm laser (3.5 mW) and normalized respect the intensity of star labelled $P_5$ peak for comparison. The peaks labelled $P_1$-$P_8$ are related to the fundamental phonon modes of the host lattice. All the modes that appear

at frequencies > 650 cm$^{-1}$ labelled with a star are consistent with the Raman active molecular modes of the corresponding guest alkylammonium species in their bromide salt.

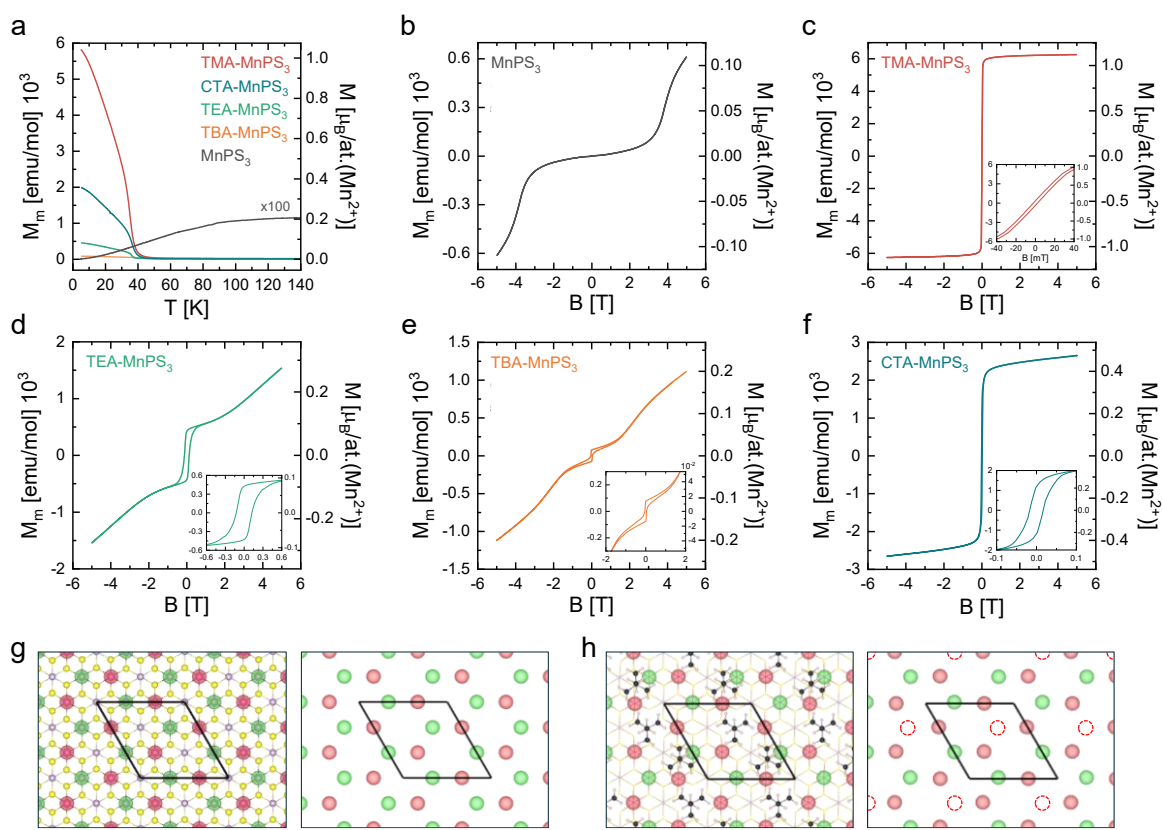

**Figure 3**. Magnetic properties of pristine and intercalated bulk crystals of MnPS$_3$. a) Temperature dependence of the magnetization measured for a pristine MnPS$_3$ crystal and for the four R$_4$N-MnPS$_3$ intercalates, under an out-of-plane magnetic field H = 0.1 T. The magnetization of the pristine MnPS$_3$ crystal has been multiplied by a factor 100. b-f) Hysteresis loops measured with the field applied out of plane for the pristine MnPS$_3$ crystal and the four intercalates. Insets in (c-f) display the low field region, characterized by the opening of a magnetic hysteresis. g-h) Spin polarized charge densities (r =r$_-$ -r$^-$) for pristine MnPS$_3$ and single Mn-vacant TMA-MnPS$_3$, respectively. Red and green colours are for up and down spin states, respectively. Density-only figures are also depicted for the sake of clarity. Mn$^{2+}$ vacancies are shown as dotted red circles.

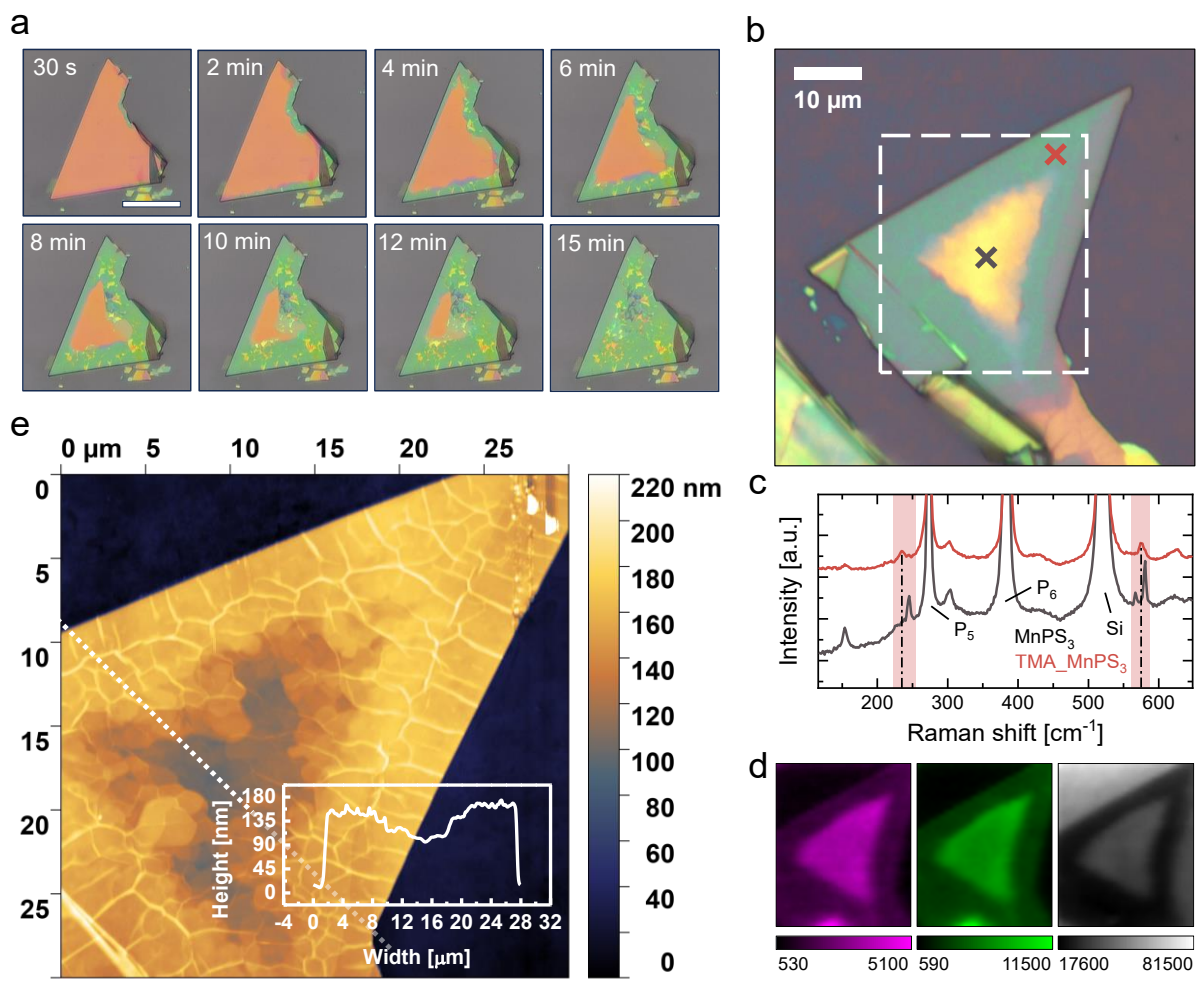

**Figure 4.** Intercalation of micromecanically exfoliated MnPS$_3$ flakes. a) Optical images displaying the process of TMA$^+$ intercalation into a selected MnPS$_3$ flake. Scale bar: b) Optical image of a partially intercalated flake, in which the intercalation has been interrupted before completion. c) Raman spectra recorded in the spots highlighted with crosses in (b). d) Raman maps displaying the intensity of P5 and P6, as well as of the Si substrate. The pristine and intercalated regions are easily distinguishable. e) Morphology of the same flake shown in (b), as measured by atomic force microscopy.

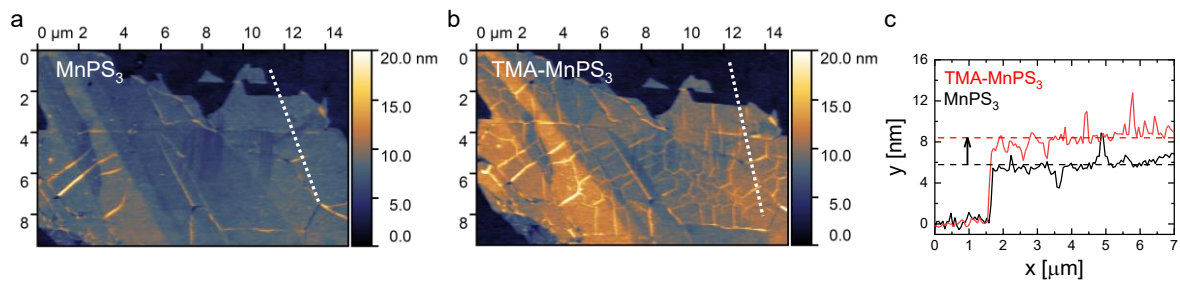

**Figure 5.** Intercalation of a few-layer-thick MnPS$_3$ flake. (a,b) Morphology of a thin flake measured before (a) and after (b) intercalation through atomic force microscopy. (c) Profile measured along the dashed lines in (a) and (b).

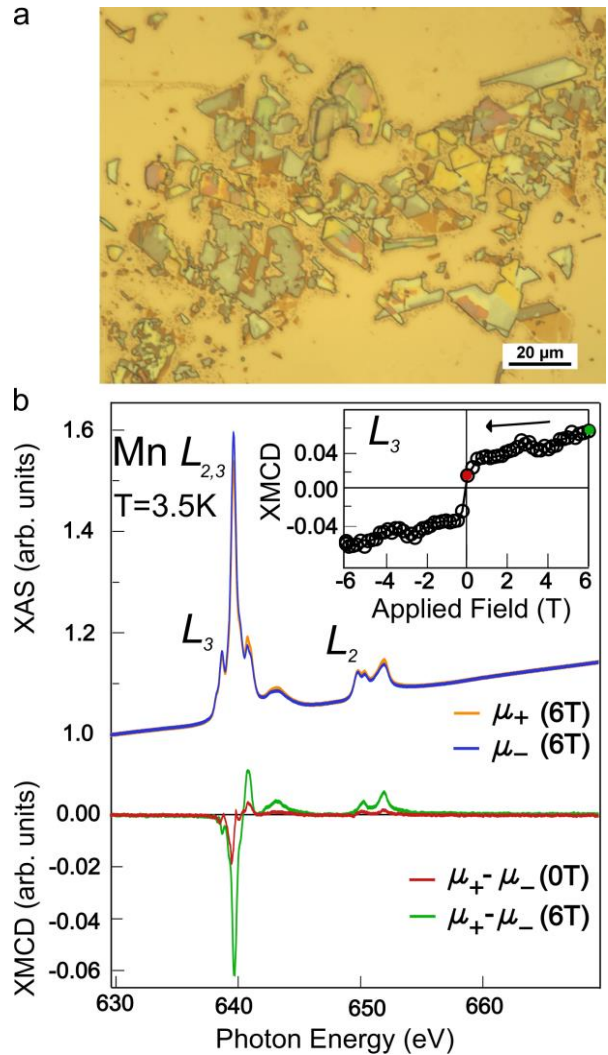

**Figure 6**. Magnetic characterization of MnPS$_3$ flakes. a) Optical image of the intercalated TMA$^+$-intercalated flakes deposited on a gold substrate. b) Top: Mn L$_{2,3}$ XAS for right (orange) and left (blue) circularly polarized light measured at 3.5 K and 6 T magnetic field. The magnetic field was applied in the direction of the X-rays and perpendicular to the investigated sample (normal incidence configuration). Inset: Hysteresis loop of the XMCD intensities at the Mn L$_3$ edge. Bottom: XMCD signal ($\mu_+ - \mu_-$) measured at 6T applied field (green) and after field removal (red curve).

**Experimental section**

*Materials*: MnPS$_3$ crystals were purchased from HQgraphene. Tetramethylammonium bromide (TMAB – purity ≥ 98%), tetraethylammonium bromide (TEAB – purity = 98%), tetrabutylammonium bromide (TBAB – purity ≥ 98%) were purchased at Aldrich and cetyltrimethylammonium bromide (CTAB – purity ≥ 99%) was purchased at Across Organics. Milli-Q water is used to prepare the solutions.

MnPS3 crystals were prepred by chemical vapor transport using iodine as a transport medium. The elements (Mn 99.95%, -100 mesh, Mateck, Germany; P 99.9999%, 2-6 mm, Wuhan Xinrong New Materials Co., China; S 99.9999%, 2-6 mm, Wuhan Xinrong New Materials, Co., China) corresponding to 30 g of MnPS3 were placed in quartz ampoule (50 x 250 mm, 3mm wall thickness) and melt sealed under high vacuum (under 1x10-3 Pa) using silicon oil diffusion pump and LN2 cooled cold trap. As a transport medium was used iodine (0.5 g, 99.9%, Fisher Scientific, USA) together with 1 at.% excess of sulfur and phosphorus above stoichiometry). The ampoule was first placed in muffle furnace to form bulk MnPS3. The ampoule was heated on 450 °C for 50 hours, on 500 °C for 50 hours and on 600 °C for 50 hours. The heating and cooling rate were 0.5 °C per minute. The formed polycrystalline MnPS3 was subsequently placed in two zone horizontal furnace for CVT crystal growth. In first stage the growth zone was heated on 750 °C and source zone on 600 °C to clean the growth area by reverse transport. After 2 days the thermal gradient was reversed and the source zone was kept at 750°C, while the growth zone was kept at 650 °C. After 10 days the ampoule was cooled on room temperature and open in argon filled glovebox, where was crystals stored before further use.

*Intercalation of crystals and flakes*: MnPS$_3$ bulk crystals (4-6 mm$^2$; 0.2 – 2 mg) were sealed in an Eppendorf vial with 1.5 mL of the aqueous solution of the different alkylammonium bromide salts. When higher temperatures were needed to accelerate the intercalation process, the Eppendorf was closed tightly with parafilm to prevent water evaporation and kept in a Thermomixer. Different solution concentration, temperature and time have been optimized to ensure a complete intercalation for each alkylammonium cation. The optimized parameters are summarized in Supporting table 1. For the intercalation of MnSP$_3$ flakes, MnPS$_3$ bulk crystals were mechanically exfoliated using scotch tape to produce flakes which were transfered directly on a Si/SiO$_2$ substrate. This substrate was immersed in a TMAB acqueous solution (1M) for desidered amount of time, rinsed many times with milliQ H$_2$O and dried in air naturally. For the Raman study of partially intercalated flakes, 3 minutes were set for the intercalation, which was stopped when the flakes were rinsed with fresh Milli-Q H$_2$O. For the XMCD experiments, the flakes were transferred on a 50-nm-thick Au film coated Si/SiO$_2$ substrate. The Au film was used as a conductive layer to ground the flakes.

*Induced Inductively coupled plasma atomic emission spectroscopy (ICP-AES)*: Elemental composition measurements were performed for the pristine MnPS$_3$ and its four alkylammonium intercalates, using an optical emission spectrophotometer with inductively coupled plasma (Agilent 5100). Pristine and intercalated MnPS3 crystals (6-8 mg for each compound) were digested in 2 mL of HNO$_3$ 69% at room temperature and diluted with milli-Q water to obtain a concentration of 1%. Concentration of analytes (Mn, P) was within the 0.01 – 10 ppm calibration range. The instrument is equipped with a quartz seaspray concentric nebulizer and a Scott-type spray chamber. The operating and acquisition parameters are the following: RF power = 1300 W, plasma gas flow = 12 L/min, auxiliar gas flow = 1.0 L/min, nebulizer gas flow = 0.7 L/min, sample flow = 1.0 mL/min, acquisition integration time = 8000 ms. The analytical detection are based on the following emission wavelength of the elements of interest: Mn (257.610 nm, 259.372 nm, 260.568 nm), P (213.618 nm, 214.914 nm).

*X-ray Diffractometry (XRD)*: XRD measurements were carried out using a Empyrean diffractometer (PANalytical) on bulk crystals and on exfoliated flakes supported on a Si/SiO$_2$ substrate. A copper cathode was used as X-ray source. Both the wavelengths K$\alpha_1$ (1.5406 Å) and K$\alpha_2$ (1.5443 Å) were employed to maximize the intensity of the diffracted beam.

*Micro-Raman spectroscopy*: Raman spectroscopy measurements were performed with a Renishaw inVia Qontor system at room temperature using a 532 nm laser and a Nikon 50× objective with long working distance (NA = 0.5). For bulk crystals, measurements were performed in atmospheric conditions with a laser power of ~ 3.5 mW. The Raman map of the partially TMA$^+$ intercalated MnPS$_3$ flake was acquired using a Linkam chamber, in vacuum conditions (~ 10$^{-6}$ mbar) at RT (see Figure S6, Supporting information). This allowed the usage of a power intensity as high as ~ 1.7 mW not damaging the sample and minimizing the total time required for the measurement. Each spectrum was collected with an integration time of 30 s. Measurements were separated by 1-μm spacing.

*Vibrating sample magnetometry (VSM)*: Magnetization vs. temperature or field H measurements were carried out using a physical properties measurement system (PPMS) in vibrating sample magnetometer (VSM) mode.

*Thermogravimetric analysis (TGA)*: The TGA Q500 from TA instruments was employed to study the thermal stability of the alkylammonium intercalated MnPS$_3$ crystal. For each intercalate, a 2-3 mg sample of crystals *wa*s loaded in a Platinum pan. The TGA chamber was continuously purged with a 60 mL/min N$_2$ flow. The samples were heated from RT to 600 °C at a rate of 1 °C/min.

*X-ray absorption (XAS) and X-ray magnetic circular dichroism (XMCD)*: XAS and XMCD measurements were carried out at BOREAS beamline of the ALBA synchrotron radiation center in Barcelona (Spain). The applied magnetic field was parallel to the light propagation vector. The XAS measurements were taken in normal incidence configuration (sample surface normal to light and field direction) in total electron yield mode with both fully (99%) left- and right-circularly polarized light. The sample signal was normalized to the incident photon flux measured as the total electron yield signal of a conductive-diamond membrane placed downstream the last optical element of the beamline. Magnetization curves were obtained by measuring the Mn $L_3$ XMCD intensity as a function of the continuously varying applied magnetic field.

For this experiment, $MnPS_3$ flakes were exfoliated onto a 50-nm-thick Au film deposited on a $Si/SiO_2$ substrate and then intercalated with $TMA^+$ molecules. The synchrotron light is defocused to approx. 1x1 $mm^2$ to ensure low photon densities and prevent beam damage. Therefore, our XMCD measurements probe the magnetic response of a collection of randomly distributed flakes, with thicknesses typically ranging from 10 nm to 100 nm.

*Atomic Force Microscopy (AFM)*: The morphological image of the $TMA^+$ intercalated flakes was acquired in tapping mode using standard tips (NCHV model, 40 N/m, 320 KHz, 8 nm tip radius) in an Agilent 5500 SPM in ambient conditions. The AFM image was taken with a scanning speed is 0.8 lines/min. and has a resolution of 1024 lines per side.

Wavelength dispersive X-ray fluorescence: Samples were analysed using a wavelength dispersive XRF (WD-XRF) spectrometer (PANalytical Axios) with Rhodium tube and three detectors (gas flow, scintillation and Xe-sealed). A Fundamental Parameters (FP) based qualitative/semiquantitative software was used, in order to compare samples and determine the absence/presence of bromine. Each sample (set of crystals) was put between films of *mylar* so that they could be analysed in vacuum.

*Optical Microscopy*: An optical microscope was used to monitor the intercalation in flakes. The $Si/SiO_2$ substrate covered with flakes was mounted onto a home-made microfluidic chamber equipped with an optical port close to the sample surface. The videos were recorded with a 20× objective and sped up for a factor 30× (Supporting Video 1), and 20× (Supporting Video 2).

The field of view in Supporting Video 1 is 58 μm × 66 μm; in Supporting Video 2 it is 87 μm × 65 μm.

*DFT calculations*: Structural and electronic optimizations of MnPS$_3$ crystals and intercalating molecules were performed using Vienna ab initio Simulation Package (VASP) software, which performs density functional-based calculations[50,51]. The generalized gradient approximation (GGA) form of Perdew-Burke-Ernzerhof (PBE) was used for the exchange-correlation functional[52]. The kinetic energy cutoff of the plane-wave basis set was taken to be 500 eV. The convergence criterion between consequent electronic and ionic steps was set at $10^{-5}$ and $10^{-4}$ eV, respectively. At least 12 Å of vacuum spacing between layers was taken in order to hinder interactions between adjacent layers.

*Statistical Analysis:* For each alkylammonium ion, multiple crystals were intercalated (typically 4, see Supporting Table S3). Moreover, we characterized the composition and magnetic properties of two batches of samples, using MnPS$_3$ cystals from different sources and fresh solutions (see Supporting Table S2 and Supporting Figure S4). The XRD patterns were normalized with respect to the intensity of (001) peaks for comparison. Several Raman spectra of intercalated crystals and flakes were obtained for each samples (10-20 measurements). The displayed spectra are representative of each samples which show homogeneity at a macroscopic scale. Stoichiometry determined by ICP-AES and reported in the manuscript is an everage of three independent measurements for each sample. Magnetic measurements were performed for two different samples for batch of intercalated crystals.

**Supporting Information**

Supporting Information is available from the Wiley Online Library or from the author.

**Acknowledgements**

This work was supported under Projects PID2021-128004NB-C21 and PID2021-122511OB-I00 and under the María de Maeztu Units of Excellence Programme (Grant CEX2020-001038-M) funded by MICIU/AEI/10.13039/501100011033 and ERDF/eu. This work was also supported by the FLAG-ERA grant MULTISPIN, with grant number PCI2021-122038-2A funded by the MICIU/AEI and European Union NextGenerationEU/PRTR and with Grant No. 221N401 funded by TUBITAK. B. M.-G. and M. G. acknowledge support from the "Ramón y Cajal" Programme (grant no. RYC2021-034836-I and RYC2021-031705-I funded by MICIU/AEI and European Union NextGenerationEU/PRTR). F. S. acknowledges the diputación foral de Gipuzkoa through Gipuzkoa Next 2023. A. A. acknowledges the Eusko Jaurlaritza (Basque Government) through Grant IT-1566-22. The XAS/XMCD experiments were performed at BOREAS beamline of the ALBA Synchrotron with the collaboration of ALBA staff (proposal number 2022097120). Computational resources were provided by TUBITAK ULAKBIM, High Performance and Grid Computing Center (TR-Grid e-Infrastructure). The authors thank SGIker Medidas Magneticas Gipuzkoa (UPV/EHU/ERDF, EU) for technical and human support. The authors also acknowledge the SGIker (UPV/EHU) SCAB for the ICP-AES measurements. Z.S. was supported by was supported by ERC-CZ program (project LL2101) from Ministry of Education Youth and Sports (MEYS) and by the project Advanced Functional Nanorobots (reg. No. CZ.02.1.01/0.0/0.0/15_003/0000444 financed by the EFRR).